\documentclass
[floatfix,superscriptaddress,secnumarabic,amssymb,amsmath,nobibnotes,aps,prd,showkeys,showpacs,nofootinbib,onecolumn,notitlepage,12pt]{revtex4}%
\usepackage{setspace}
\usepackage{color}
\usepackage{amsmath}
\usepackage{amsfonts}
\usepackage{verbatim}
\usepackage{amssymb}
\usepackage{graphicx,bm}
\usepackage{graphicx}
\usepackage{amsmath}
\usepackage{amssymb}
\usepackage{amssymb}
\usepackage{graphicx,bm}
\usepackage{graphicx}
\usepackage[caption=false]{subfig}%
\providecommand{\U}[1]{\protect\rule{.1in}{.1in}}

\newcommand{\be}{\begin{equation}}
\newcommand{\ee}{\end{equation}}

\newcommand{\mincir}{\raise
-3.truept\hbox{\rlap{\hbox{$\sim$}}\raise4.truept\hbox{$<$}\ }}
\newcommand{\magcir}{\raise
-3.truept\hbox{\rlap{\hbox{$\sim$}}\raise4.truept\hbox{$>$}\ }}

\ifx\pdfoutput\relax\let\pdfoutput=\undefined\fi
\newcount\msipdfoutput
\ifx\pdfoutput\undefined\else
\ifcase\pdfoutput\else
\ifx\paperwidth\undefined\else
\ifdim\paperheight=0pt\relax\else\pdfpageheight\paperheight\fi
\ifdim\paperwidth=0pt\relax\else\pdfpagewidth\paperwidth\fi
\fi\fi\fi
\begin{document}
\title{Symmetric teleparallel cosmology with boundary corrections}
\author{Andronikos Paliathanasis}
\email{anpaliat@phys.uoa.gr}
\affiliation{Institute of Systems Science, Durban University of Technology, Durban 4000,
South Africa}
\affiliation{Departamento de Matem\'{a}ticas, Universidad Cat\'{o}lica del Norte, Avda.
Angamos 0610, Casilla 1280 Antofagasta, Chile}

\begin{abstract}
We investigate the geometrodynamical effects of introducing the boundary term
in symmetric teleparallel gravity. Specifically, we consider a homogeneous and
isotropic universe in $f\left(  Q, B \right)  $, where $Q$ is the
non-metricity scalar, and $B$ is the boundary term that relates the
non-metricity and Ricci scalars. For the connection in the coincidence gauge,
we find that the field equations are of fourth-order, and the fluid components
introduced by the boundary are attributed to a scalar field. In the
coincidence gauge, the cosmological field equations are equivalent to those of
teleparallelism with a boundary term. Nevertheless, for the connection defined
in the non-coincidence gauge, the geometrodynamical fluid consists of three
scalar fields. We focus on the special case of $f\left(  Q, B \right)  = Q +
F\left(  B \right)  $ theory, and we determine a new analytic cosmological
solution that can explain the late-time acceleration of the universe and
provide a geometric mechanism for the unification of dark energy with dark matter.

\end{abstract}
\maketitle

\section{Introduction}

\label{sec1}

General Relativity is a well-tested theory \cite{gw1}; however, it faces
challenges on cosmological scales \cite{ten1,ten1a,ten2}. Recent cosmological
observations \cite{rr1, Teg, Kowal, Komatsu, suzuki11} suggest that the
universe is currently experiencing an acceleration phase driven by an exotic
matter source known as dark energy \cite{jo1, jo}. On the other hand, to
address the homogeneity and flatness problems, it has been proposed that the
universe underwent an inflationary epoch in its early stages driven by the
inflaton field \cite{inf1}.

The nature and physical properties of the inflaton and dark energy remain
subjects of debate among cosmologists. Cosmologists have proposed different
models to explain cosmic acceleration, which can be broadly categorized into
two groups. In the first category, a matter source with negative pressure is
introduced into the field equations of General Relativity \cite{ratra, de1,
de2, de3, de4, de5}. In the second category of models, cosmologists focus on
determining a new gravitational theory by modifying the gravitational Action
Integral \cite{de01, de02, de03}.

The consideration of quantum-gravitational effects in the one-loop
approximation in gravity \cite{df1, df2} leads to the introduction of the
$R^{2}$ term in the gravitational Lagrangian. This quadratic theory of gravity
\cite{qua1, qua2, qua3} has been used as a mechanism to explain inflation
\cite{star, barc, stt1}. Indeed, when the $R^{2}$ term dominates, it leads to
a de Sitter expansion \cite{bar1}. The success of the $R^{2}$ term in
describing acceleration has given rise to a family of theories known as
$f(R)$-gravity \cite{Buda}. For a comprehensive review, we recommend referring
to \cite{Sotiriou}.

The symmetric teleparallel $f(Q)$-theory of gravity has garnered the attention
of cosmologists \cite{lav2, lav3}. $f(Q)$-theory is defined within the
framework of symmetric teleparallel theory \cite{nester, nester1}, where the
physical space is described by the metric tensor. However, the fundamental
connection in this theory is not the Levi-Civita connection but a symmetric
and flat connection, leading to the definition of a nonzero non-metricity
scalar $Q$. When $f(Q)$ is linear, the gravitational theory becomes equivalent
to General Relativity, known as Symmetric Teleparallel General Relativity
(STGR) \cite{nester}. In STGR, the Ricci scalar $R$ and the non-metricity
scalar $Q$ differ only by a boundary term $B$, which means that the variation
of the boundary term is neglected, resulting in equivalent field equations.
There is a plethora of studies in the literature on $f(Q)$-theory; for
example, see \cite{fq1, fq2, fq3, fq4, fq5, fq6, fq7, fq8, fq9,fq10,fq11}, and
references therein.

In the process of constructing a gravitational theory, the boundary term $B$
has been utilized to modify the gravitational Action Integral. Inspired by
teleparallelism \cite{bah}, the symmetric teleparallel theory with a boundary
term, expressed as $f(Q,B)$-gravity, was introduced in \cite{ftc1, ftc2}. It
was discovered that the introduction of the boundary term further modifies the
field equations of the gravitational model. In this study, we aim to
investigate the effects of the boundary term on cosmological dynamics.

Because the definition of the symmetric and flat connection is not unique,
there can be multiple different connections that describe the same
gravitational model. The corresponding non-metricity constants differ by a
boundary term, which means that they lead to the same equations in STGR.
However, this is not true in the case of a nonlinear Lagrangian. As a result,
the definition of the connection affects the boundary term, implying that
$f(Q,B)$-gravity depends on the definition of the connection.

Recently, in \cite{min}, it was proven that $f(Q)$-theory of gravity admits a
minisuperspace description, and the theory is of fourth- or sixth-order,
depending on the connection used. The higher-order derivatives can be
attributed to a scalar field, and the gravitational Lagrangian can be written
as that of a higher-dimensional second-order dynamical system. This approach
is applied in this study in the context of $f(Q,B)$-gravity for a homogeneous
and isotropic geometry. The structure of the paper is as follows.

In Section \ref{sec2}, we present the definition of symmetric teleparallel
gravity and its generalization, the $f(Q)$-theory, where $Q$ represents the
non-metricity scalar. The extension of symmetric teleparallel theory with a
boundary term $B$ is discussed in Section \ref{sec3}. For the $f(Q,B)$-theory,
we discuss when it is equivalent to teleparallel $f(T,B_{T})$ theory and when
it approaches the limit of $f(R)$-gravity. In Section \ref{sec4}, we present
the gravitational field equations for a homogeneous and isotropic universe.
Specifically, we employ Lagrange multipliers to demonstrate that the field
equations in $f(Q,B)$-theory can be described within a minisuperspace
framework. It's important to note that the definition of the symmetric and
flat connection in a FLRW universe is not unique. Therefore, we provide the
field equations for all four different families of connections. One novel
aspect of the Lagrange multiplier approach is the introduction of scalar
fields to account for the higher-order derivatives in the gravitational model.
Consequently, $f(Q,B)$-cosmology is a fourth-order gravitational theory,
equivalent to teleparallel $f(T,B_{T})$-cosmology, when the symmetric and flat
connection is defined in the coincidence gauge. However, when the connection
is defined in the non-coincidence gauge, the field equations become
eighth-order and are described by three scalar fields.

Furthermore, in Section \ref{sec5}, we focus on the particular case of $f(Q,B)
= Q + F(B)$ cosmology, where the gravitational Action Integral is modified by
nonlinear terms of the boundary. For this specific gravitational model, the
order of the gravitational field equations is reduced by two in the
non-coincidence gauge. Consequently, the geometric fluid is described by two
scalar fields. In Section \ref{sec6}, we present a new analytic solution for
symmetric teleparallel cosmology with a boundary term in the non-coincidence
gauge. This new solution is capable of describing the $\Lambda$CDM universe at
the present time and possesses a de Sitter attractor. Finally, in Section
\ref{sec7}, we summarize our findings and draw our conclusions.

\section{Symmetric teleparallel gravity}

\label{sec2}

We examine a gravitational model described by the four-dimensional metric
tensor $g_{\mu\nu}$ and the covariant derivative $\nabla_{\lambda}~$ which is
defined using the generic connection $\Gamma_{\mu\nu}^{\kappa}\,$, such that
the autoparallels are defined as \cite{auto}%
\[
\frac{d^{2}x^{\mu}}{ds^{2}}+\Gamma_{\kappa\nu}^{\mu}\frac{dx^{\kappa}}%
{ds}\frac{dx^{\nu}}{ds}=0.
\]
Connection~ $\Gamma_{\mu\nu}^{\kappa}$ determines the nature of the geometry.

For the general connection we can define the the Riemann tensor%
\begin{equation}
R_{\;\lambda\mu\nu}^{\kappa}=\frac{\partial\Gamma_{\;\lambda\nu}^{\kappa}%
}{\partial x^{\mu}}-\frac{\partial\Gamma_{\;\lambda\mu}^{\kappa}}{\partial
x^{\nu}}+\Gamma_{\;\lambda\nu}^{\sigma}\Gamma_{\;\mu\sigma}^{\kappa}%
-\Gamma_{\;\lambda\mu}^{\sigma}\Gamma_{\;\mu\sigma}^{\kappa},
\end{equation}
the torsion tensor%
\begin{equation}
\mathrm{T}_{\mu\nu}^{\lambda}=\Gamma_{\;\mu\nu}^{\lambda}-\Gamma_{\;\nu\mu
}^{\lambda},
\end{equation}
and the non-metricity tensor~\cite{Eisenhart}
\[
Q_{\lambda\mu\nu}=\nabla_{\lambda}g_{\mu\nu}=\frac{\partial g_{\mu\nu}%
}{\partial x^{\lambda}}-\Gamma_{\;\lambda\mu}^{\sigma}g_{\sigma\nu}%
-\Gamma_{\;\lambda\nu}^{\sigma}g_{\mu\sigma}.
\]

In General Relativity, $\Gamma_{\mu\nu}^{\kappa}$ is recognized as the
Levi-Civita connection, denoted as $\tilde{\Gamma}{\mu\nu}^{\kappa}$.
Consequently, in this framework, $\mathrm{T}{\mu\nu}^{\lambda}=0$ and
$Q_{\lambda\mu\nu}=0$. Therefore, the primary scalar in General Relativity is
the Ricci scalar $R$.

On the other hand, in the Teleparallel Equivalent of General Relativity (TEGR)
\cite{de03}, the connection $\Gamma_{\mu\nu}^{\kappa}$ is replaced by the
antisymmetric Weitzenb"{o}ck connection, resulting in $R_{;\lambda\mu\nu
}^{\kappa}=0$ and $Q_{\lambda\mu\nu}=0$. In this context, the torsion scalar
$T$ takes on the role of the fundamental geometric object in teleparallel gravity.

In the theory under consideration, which is STGR, $\Gamma_{\mu\nu}^{\kappa}$
possesses the property of being both flat and torsionless. This implies that
$R_{;\lambda\mu\nu}^{\kappa}=0$ and $\mathrm{T}{\mu\nu}^{\lambda}=0$.
Additionally, it inherits the symmetries of the metric tensor $g{\mu\nu}$.
Thus, the nonmetricity scalar $Q$ defined as \cite{nester}
\begin{equation}
Q=Q_{\lambda\mu\nu}P^{\lambda\mu\nu}%
\end{equation}
is the fundamental geometric quantity of gravity.

Tensor $P^{\lambda\mu\nu}$ is defined as \cite{nester}%
\begin{equation}
P_{\;\mu\nu}^{\lambda}=-\frac{1}{4}Q_{\;\mu\nu}^{\lambda}+\frac{1}{2}%
Q_{(\mu\phantom{\lambda}\nu)}^{\phantom{(\mu}\lambda\phantom{\nu)}}+\frac
{1}{4}\left(  Q^{\lambda}-\bar{Q}^{\lambda}\right)  g_{\mu\nu}-\frac{1}%
{4}\delta_{\;(\mu}^{\lambda}Q_{\nu)}, \label{defP}%
\end{equation}
which is written with the help of the traces\footnote{Parenthesis in the
indices denote symmetrization, that is, $A_{(\mu\nu)}=\frac{1}{2}\left(
A_{\mu\nu}+A_{\nu\mu}\right)  $; and $\delta_{\;\nu}^{\mu}$ is the Kroncker
delta.} $Q_{\mu}=Q_{\mu\nu}^{\phantom{\mu\nu}\nu}$ and $\bar{Q}_{\mu
}=Q_{\phantom{\nu}\mu\nu}^{\nu\phantom{\mu}\phantom{\mu}}$.

The Ricci scalar $R$ correspond to the Levi-Civita connection $\tilde{\Gamma
}_{\mu\nu}^{\kappa}$ of the metric tensor $g_{\mu\nu}$, and the nonmetricity
scalar $Q$ for a symmetric and flat connection $\Gamma_{\mu\nu}^{\kappa}$
differ by a boundary term $B,$defined as $B=R-Q~$.

The gravitational Action Integral of STGR reads \cite{nester}
\begin{equation}
\int d^{4}x\sqrt{-g}Q\simeq\int d^{4}x\sqrt{-g}R+\text{boundary terms.}%
\end{equation}
from where it follows that STGR is dynamically equivalent to GR.

However, when nonlinear components of the non-metricity scalar $Q$ are
introduced in the gravitational Action, as in $f(Q)$-gravity, this equivalence
is lost. Nevertheless, the corresponding gravitational theory does not possess
any dynamical equivalence with General Relativity or its generalization,
$f(R)$-gravity.

The Action Integral in symmetric teleparallel $f(Q)$-gravity is defined
\cite{lav2,lav3}
\begin{equation}
S_{f\left(  Q\right)  }=\int d^{4}x\sqrt{-g}f(Q).
\end{equation}
The resulting field equations are
\begin{equation}
\frac{2}{\sqrt{-g}}\nabla_{\lambda}\left(  \sqrt{-g}f_{,Q}P_{\;\mu\nu
}^{\lambda}\right)  -\frac{1}{2}f(Q)g_{\mu\nu}+f_{,Q}\left(  P_{\mu\rho\sigma
}Q_{\nu}^{\;\rho\sigma}-2Q_{\rho\sigma\mu}P_{\phantom{\rho\sigma}\nu}%
^{\rho\sigma}\right)  =0, \label{fl1}%
\end{equation}%
\begin{equation}
\nabla_{\mu}\nabla_{\nu}\left(  \sqrt{-g}f_{,Q}P_{\phantom{\mu\nu}\sigma}%
^{\mu\nu}\right)  =0. \label{kl.01}%
\end{equation}

Equations (\ref{fl1}) represent the modified Einstein field equations in
$f(Q)$-gravity, while equation (\ref{kl.01}) corresponds to the equation of
motion for the connection. . When equation (\ref{kl.01}) holds true for a
specific connection at all times, that connection is referred to as the
\textquotedblleft coincidence gauge\textquotedblright. However, if the
equation (\ref{kl.01}) is not always satisfied for a particular connection,
then that connection is defined within the so-called \textquotedblleft
non-coincidence gauge\textquotedblright as discussed in \cite{lav3}.

\section{Symmetric teleparallel boundary gravity}

\label{sec3}

Recently, a generalization of the $f(Q)$ theory was introduced \cite{ftc1,
ftc2} by incorporating a boundary term into the gravitational Action Integral.
More precisely, this extended framework includes the gravitational Action
Integral as follows
\begin{equation}
S_{f\left(  Q,B\right)  }=\int d^{4}x\sqrt{-g}f(Q,B)~ \label{ftc}%
\end{equation}
where $B=R-Q$.

In the notation of \cite{ftc2}, the boundary term is noted as $C$. This
gravitational theory is equivalent to General Relativity, when $f\left(
Q,B\right)  $ is a linear function, that is, $f\left(  Q,B\right)
=f_{1}Q+f_{2}B-2\Lambda$; the limit of $f\left(  Q\right)  $ is recovered when
$f\left(  Q,B\right)  =f\left(  Q\right)  +f_{2}B$; while the fourth-order
$f\left(  R\right)  $ theory of gravity is recovered when $f\left(
Q,B\right)  =f\left(  Q+B\right)  $.

The gravitational field equations which correspond to the Action Integral
(\ref{ftc}) are
\begin{align}
0  &  =\frac{2}{\sqrt{-g}}\nabla_{\lambda}\left(  \sqrt{-g}f_{,Q}P_{\;\mu\nu
}^{\lambda}\right)  -\frac{1}{2}f(Q,B)g_{\mu\nu}+f_{,Q}\left(  P_{\mu
\rho\sigma}Q_{\nu}^{\;\rho\sigma}-2Q_{\rho\sigma\mu}P_{\phantom{\rho\sigma}\nu
}^{\rho\sigma}\right) \nonumber\\
&  +\left(  \frac{B}{2}g_{\mu\nu}-\nabla_{\mu}\nabla_{\nu}+g_{\mu\nu}%
g^{\kappa\lambda}\nabla_{\kappa}\nabla_{\lambda}-2P_{~~\mu\nu}^{\lambda}%
\nabla_{\lambda}\right)  f_{,B}%
\end{align}

$f\left(  Q,B\right)  $ has been inspired by the teleparallel boundary gravity
$f\left(  T,B_{T}\right)  ~$\cite{bah}, where now $B_{T}$ is the boundary term
relates the Ricciscalar~$R$ and the torsion scalar $T$ for the Weitzenb\"{o}ck
connection \cite{Weitzenb23}. In a similar way, $f\left(  T,B_{T}\right)  $
recovers $f\left(  R\right)  $-gravity when $f\left(  T,B_{T}\right)
=f\left(  T+B_{T}\right)  ~$\cite{bah}, while the limit of GR is recovered
when $f\left(  T,B_{T}\right)  =f_{1}T+f_{2}B_{T}-2\Lambda;$ and $f\left(
T\right)  $-theory follows when $f\left(  T,B_{T}\right)  =f\left(  T\right)
+f_{2}B_{T}$. There are many astrophysical and cosmological applications of
$f\left(  T,B_{T}\right)  $-theory in the literature
\cite{ftb1,ftb2,ftb3,ftb4,ftb5,ftb6}. From these results it is clear that the
boundary $B_{T}$ plays an important role in geometric description of dark
energy. Thus, the generalization if symmetric teleparallel theory seems natural.

$f\left(  T,B_{T}\right)  $ theory of gravity is a fourth-order theory,
similar to $f\left(  R\right)  $-theory, while when $f\left(  T,B_{T}\right)
=f\left(  T\right)  +f_{2}B_{T}$ the resulting field equations are of
second-order \cite{ftb7}. The order of symmetric teleparallel $f\left(
Q\right)  $-theory depends on the connection which is used for the definition
of the nonmetricity scalar $Q$.\ For the connection defined in the coincidence
gauge $f\left(  Q\right)  $ is a second-order theory, while for a connection
in the non-coincidence gauge $f\left(  Q\right)  $-theory is a sixth-order
theory of gravity.

To comprehend the degrees of freedom within $f(Q,B)$-theory, we turn our
attention to a cosmological model representing an isotropic and homogeneous
universe. Employing the Lagrange multipliers method, we introduce scalar
fields that account for the dynamical degrees of freedom within the
$f(Q,B)$-theory. Consequently, our analysis reveals that the theory introduces
either one or three scalar fields.

\section{Isotropic and Homogeneous Universe}

\label{sec4}

The isotropic and homogeneous universe is described by the FLRW line element
\begin{equation}
ds^{2}=-N(t)^{2}dt^{2}+a(t)^{2}\left[  \frac{dr^{2}}{1-kr^{2}}+r^{2}\left(
d\theta^{2}+\sin^{2}\theta d\varphi^{2}\right)  \right]  , \label{genlineel}%
\end{equation}
in which $N\left(  t\right)  $ is the lapse function and $a\left(  t\right)  $
is the scale factor denotes the radius of the universe. Hence, $H=\frac{1}%
{N}\frac{\dot{a}}{a}$, where $\dot{a}=\frac{da}{dt}$ is the Hubble function.
Parameter $k$ is the spatial curvature, for $k=0$, the universe is spatially
flat, $k=+1$ corresponds to a closed FLRW\ geometry and $k=-1$ describes an
open universe.

FLRW spacetime admits a sixth-dimensional Killing algebra consisted by the
vector fields%
\begin{equation}
\zeta_{1}=\sin\varphi\partial_{\theta}+\frac{\cos\varphi}{\tan\theta}%
\partial_{\varphi},\quad\zeta_{2}=-\cos\varphi\partial_{\theta}+\frac
{\sin\varphi}{\tan\theta}\partial_{\varphi},\quad\zeta_{3}=-\partial_{\varphi}
\label{Kil1}%
\end{equation}
\begin{equation}%
\begin{split}
\xi_{1}  &  =\sqrt{1-kr^{2}}\sin\theta\cos\varphi\partial_{r}+\frac
{\sqrt{1-kr^{2}}}{r}\cos\theta\cos\varphi\partial_{\theta}-\frac
{\sqrt{1-kr^{2}}}{r}\frac{\sin\varphi}{\sin\theta}\partial_{\varphi}\\
\xi_{2}  &  =\sqrt{1-kr^{2}}\sin\theta\sin\varphi\partial_{r}+\frac
{\sqrt{1-kr^{2}}}{r}\cos\theta\sin\varphi\partial_{\theta}+\frac
{\sqrt{1-kr^{2}}}{r}\frac{\cos\varphi}{\sin\theta}\partial_{\varphi}\\
\xi_{3}  &  =\sqrt{1-kr^{2}}\cos\theta\partial_{r}-\frac{\sqrt{1-kr^{2}}}%
{r}\sin\theta\partial_{\varphi}.
\end{split}
\label{Kil2}%
\end{equation}

\subsection{Non-zero spatial curvature $k\neq0$}

For the FLRW with $k\neq0$, there exist a unique connection defined in the
non-coincidence gauge with non-zero components \cite{Heis2,Zhao}%
\begin{equation}%
\begin{split}
&  \Gamma_{\;tr}^{r}=\Gamma_{\;rt}^{r}=\Gamma_{\;t\theta}^{\theta}%
=\Gamma_{\;\theta t}^{\theta}=\Gamma_{\;t\varphi}^{\varphi}=\Gamma_{\;\varphi
t}^{\varphi}=-\frac{k}{\gamma(t)},\quad\Gamma_{\;rr}^{r}=\frac{kr}{1-kr^{2}%
},\\
&  \Gamma_{\;\theta\theta}^{r}=-r\left(  1-kr^{2}\right)  ,\quad
\Gamma_{\;\varphi\varphi}^{r}=-r\sin^{2}(\theta)\left(  1-kr^{2}\right)
\quad\Gamma_{\;r\theta}^{\theta}=\Gamma_{\;\theta r}^{\theta}=\Gamma
_{\;r\varphi}^{\varphi}=\Gamma_{\;\varphi r}^{\varphi}=\frac{1}{r},\\
&  \Gamma_{\;\varphi\varphi}^{\theta}=-\sin\theta\cos\theta,\quad
\Gamma_{\;\theta\varphi}^{\varphi}=\Gamma_{\;\varphi\theta}^{\varphi}%
=\cot\theta,
\end{split}
\end{equation}
and%
\[
\Gamma_{\;tt}^{t}=-\frac{k+\dot{\gamma}(t)}{\gamma(t)},\quad\Gamma_{\;rr}%
^{t}=\frac{\gamma(t)}{1-kr^{2}}\quad\Gamma_{\;\theta\theta}^{t}=\gamma
(t)r^{2},\quad\Gamma_{\;\varphi\varphi}^{t}=\gamma(t)r^{2}\sin^{2}(\theta).
\]

For this connection, the calculation of the non-metricity scalar is produced
\begin{equation}
Q_{k}=-6\left(  H^{2}-\frac{k}{a^{2}}\right)  +\frac{3}{a^{3}N}\left(
aN\gamma-k\frac{a^{3}}{\gamma N}\right)  ^{\cdot}. \label{bd.001}%
\end{equation}

From the Levi-Civita connection of spacetime (\ref{genlineel}) we calculate
the Ricciscalar%
\begin{equation}
R_{k}=6\left(  2H^{2}+\frac{k}{a^{2}}\right)  +\frac{6}{N}\dot{H}.
\label{bd.002}%
\end{equation}
Consequently, the boundary term$~B=R-Q~$\cite{ftc2} is
\begin{equation}
B_{k}=R_{k}-Q_{k}=3\left(  6H^{2}+\frac{2}{N}\dot{H}-\frac{1}{a^{3}N}\left(
aN\gamma-k\frac{a^{3}}{\gamma N}\right)  ^{\cdot}\right)  . \label{bd.003}%
\end{equation}

In order to derive the gravitational field equations we apply the mathematical
manipulation introduced in \cite{min} and we introduce the scalar field $\Psi$
such that $\gamma=\frac{1}{\dot{\Psi}}$.

We introduce in\ (\ref{ftc}) the Lagrange multipliers~$\lambda_{1}$ and
$\lambda_{2}$ such that%
\begin{equation}
S_{f\left(  Q,B\right)  }=\int d^{4}x\sqrt{-g}\left(  f(Q,B)~-\lambda
_{1}\left(  Q-Q_{k}\right)  -\lambda_{2}\left(  B-B_{k}\right)  \right)  .
\label{bd.01}%
\end{equation}

Variation with respect to the non-metricity scalar $Q$ and the boundary term
$B$, gives $\lambda_{1}=f_{,Q}$ and $\lambda_{2}=f_{,B}$.

Thus, by replacing in (\ref{bd.01}) it follows%

\begin{equation}
S_{f\left(  Q,B\right)  }=\int dt\left(  Na^{3}\left(  f-f_{,Q}-f_{,B}\right)
+Na^{3}f_{,Q}Q_{k}+Na^{3}f_{,B}B_{k}\right)  .
\end{equation}

Hence, integration by parts gives%
\[
\int dt\left(  Na^{3}f_{,Q}Q_{k}\right)  =\int dt\left(  -6Na^{3}f_{,Q}\left(
H^{2}-\frac{k}{a^{2}}\right)  +3f_{,Q}\left(  a\frac{N}{\dot{\Psi}}%
-k\frac{a^{3}}{N}\dot{\Psi}\right)  ^{\cdot}\right)  ,
\]

\[
\int dt\left(  Na^{3}f_{,B}B_{k}\right)  =\int dt\left(  18Na^{3}f_{,B}%
H^{2}+6a^{3}f_{,B}\dot{H}-3f_{,B}\left(  aN\gamma-k\frac{a^{3}}{\gamma
N}\right)  ^{\cdot}\right)  .
\]
Then%
\[
S_{f\left(  Q,B\right)  }=\int dt\left(
\begin{array}
[c]{c}%
Na^{3}\left(  f-f_{,Q}-f_{,B}\right)  -6\left(  f_{,Q}-3f_{,B}\right)  \left(
Na^{3}H^{2}\right) \\
+6Na^{3}f_{,Q}\frac{k}{a^{2}}+6f_{,B}a^{3}\dot{H}+3\left(  f_{,Q}%
-f_{,B}\right)  \left(  a\frac{N}{\dot{\Psi}}-k\frac{a^{3}}{N}\dot{\Psi
}\right)  ^{\cdot}%
\end{array}
\right)  .
\]
It follows%
\[
\int6f_{,B}a^{3}\dot{H}dt=\int\left(  -18Nf_{,B}a^{3}H^{2}-6\dot{f}_{,B}%
a^{3}H\right)  dt,
\]%
\[
\int3\left(  f_{,Q}-f_{,B}\right)  \left(  a\frac{N}{\dot{\Psi}}-k\frac{a^{3}%
}{N}\dot{\Psi}\right)  ^{\cdot}dt=\int-3\left(  \dot{f}_{,Q}-\dot{f}%
_{,B}\right)  \left(  a\frac{N}{\dot{\Psi}}-k\frac{a^{3}}{N}\dot{\Psi}\right)
dt.
\]

The minisuperspace Lagrangian is
\begin{align}
L\left(  N,a,\dot{a},Q,\dot{Q},B,\dot{B},\Psi,\dot{\Psi}\right)   &
=-\frac{6}{N}f_{,Q}a\dot{a}^{2}+6Naf_{,Q}k-\frac{6}{N}a^{2}\dot{f}_{,B}\dot
{a}\nonumber\\
&  -3\left(  \dot{f}_{,Q}-\dot{f}_{,B}\right)  \left(  a\frac{N}{\dot{\Psi}%
}-k\frac{a^{3}}{N}\dot{\Psi}\right)  +Na^{3}\left(  f-f_{,Q}-f_{,B}\right)
\end{align}
or equivalently%
\begin{equation}
L\left(  N,a,\dot{a},\phi,\dot{\phi},\zeta,\dot{\zeta},\Psi,\dot{\Psi}\right)
=-\frac{6}{N}\phi a\dot{a}^{2}+6Na\phi k-\frac{6}{N}a^{2}\dot{a}\dot{\zeta
}-3\left(  \dot{\phi}-\dot{\zeta}\right)  \left(  a\frac{N}{\dot{\Psi}}%
-k\frac{a^{3}}{N}\dot{\Psi}\right)  +Na^{3}V\left(  \phi,\zeta\right)  .
\label{lan.01}%
\end{equation}
in which $\phi=f_{,Q}$, $\zeta=f_{,B}$ and $V\left(  \phi,\zeta\right)
=\left(  f-f_{,Q}-f_{,B}\right)  $.

In the four-dimensional space ${a, \phi, \zeta, \Psi}$, we calculate $\left|
\frac{\partial^{2} L}{\partial q \partial q}\right|  = \frac{324a^{6}}%
{N^{4}\dot{\Psi}^{4}}(N^{2} + ka^{2}\dot{\Psi}^{2})^{2} \neq0$. This implies
that the field equations are of eighth-order and are described by the three
scalar fields $\phi, \zeta, \Psi$. The Lagrangian function (\ref{lan.01})
represents a singular dynamical system in which variation with respect to the
lapse function $N$ yields the modified Friedmann equation.

Moreover, variation with respect to the dynamical variables $\left\{
a,\phi,\zeta,\Psi\right\}  $ gives four second-order differential equations.

For a constant lapse function, i.e. $N=1$, the cosmological field equation
are
\begin{equation}
0=3\phi H^{2}+3\phi\frac{k}{a^{2}}+3H\dot{\zeta}-\frac{3}{2}\left(  \dot{\phi
}-\dot{\zeta}\right)  \left(  \frac{1}{a^{2}\dot{\Psi}}+k\dot{\Psi}\right)
+\frac{1}{2}V\left(  \phi,\zeta\right)  , \label{lan.02}%
\end{equation}%
\begin{equation}
0=\dot{H}+\frac{3}{2}H^{2}+\frac{k}{2a^{2}}+\frac{V}{4\phi}+H\frac{\dot{\phi}%
}{\phi}+\frac{\dot{\zeta}-\dot{\phi}}{\phi}\left(  \frac{1}{4a^{2}\dot{\Psi}%
}-\frac{3k}{4}\dot{\Psi}\right)  +\frac{1}{2}\ddot{\zeta}, \label{lan.03}%
\end{equation}%
\begin{equation}
0=3\left(  1+ka^{2}\dot{\Psi}^{2}\right)  \frac{\ddot{\Psi}}{\Psi}-\left(
3H+6k\dot{\Psi}-a^{2}\dot{\Psi}\left(  6H^{2}+9kH\dot{\Psi}-V_{,\phi}\right)
\right)  ,
\end{equation}%
\begin{equation}
0=6\dot{H}+9H\left(  2H+k\dot{\Psi}\right)  +3k\ddot{\Psi}+V_{,\zeta}-\frac
{3}{a^{2}\dot{\Psi}^{2}}\left(  H\dot{\Psi}-\ddot{\Psi}\right)  ,
\end{equation}%
\begin{equation}
0=3a\dot{\Psi}\left(  1+a^{2}k\dot{\Psi}^{2}\right)  \left(  \ddot{\zeta
}-\ddot{\phi}\right)  +a\left(  \dot{\zeta}-\dot{\phi}\right)  \left(
\dot{\Psi}H\left(  1+3a^{2}k\dot{\Psi}^{2}\right)  -2\ddot{\Psi}\right)  .
\end{equation}

The modified Friedmann's equations (\ref{lan.02}), (\ref{lan.03}) can be
written in the equivalent form%
\begin{align}
3\left(  H^{2}+\frac{k}{a^{2}}\right)   &  =G_{eff}\rho_{f\left(  Q,B\right)
}^{\Gamma_{k}},\\
-2\dot{H}-3H-\frac{k}{a^{2}}  &  =G_{eff}p_{f\left(  Q,B\right)  }^{\Gamma
_{k}},
\end{align}
in which $\rho_{f\left(  Q,B\right)  }$, $p_{f\left(  Q,B\right)  }$ are the
components for the geometric fluid which follows by the nonlinear $f\left(
Q,B\right)  $-theory defined as%
\begin{align}
\rho_{f\left(  Q,B\right)  }^{\Gamma_{k}}  &  =-\left(  3H\dot{\zeta}-\frac
{3}{2}\left(  \dot{\phi}-\dot{\zeta}\right)  \left(  \frac{1}{a^{2}\dot{\Psi}%
}+k\dot{\Psi}\right)  +\frac{1}{2}V\left(  \phi,\zeta\right)  \right)
,\label{lan.04}\\
p_{f\left(  Q,B\right)  }^{\Gamma_{k}}  &  =\frac{V}{2}+2H\dot{\phi}%
+\frac{\dot{\zeta}-\dot{\phi}}{2}\left(  \frac{1}{4a^{2}\dot{\Psi}}-\frac
{3k}{4}\dot{\Psi}\right)  . \label{lan.05}%
\end{align}
and%
\begin{equation}
G_{eff}=\frac{1}{\phi}\text{.}%
\end{equation}
We remark that scalar field $\phi$ is defined in the Jordan frame.

\subsection{Spatially flat case $k=0$}

For the spatially flat case, $k=0$, it has been found that there exist three
families of connections. The common non-zero coefficients of the three
connections are%
\[
\Gamma_{\theta\theta}^{r}=-r~,~\Gamma_{\varphi\varphi}^{r}=-r\sin^{2}\theta
\]%
\[
\Gamma_{\varphi\varphi}^{\theta}=-\sin\theta\cos\theta~,~\Gamma_{\theta
\varphi}^{\varphi}=\Gamma_{\varphi\theta}^{\varphi}=\cot\theta
\]%
\[
\Gamma_{\;r\theta}^{\theta}=\Gamma_{\;\theta r}^{\theta}=\Gamma_{\;r\varphi
}^{\varphi}=\Gamma_{\;\varphi r}^{\varphi}=\frac{1}{r}%
\]
while the additional components for connections $\Gamma_{1},~\Gamma_{2}$ and
$\Gamma_{3}$ are \cite{Heis2,Zhao}
\[
\Gamma_{1}:\Gamma_{\;tt}^{t}=\gamma(t),
\]%
\[
\Gamma_{2}:\Gamma_{\;tt}^{t}=\frac{\dot{\gamma}(t)}{\gamma(t)}+\gamma
(t),\quad\Gamma_{\;tr}^{r}=\Gamma_{\;rt}^{r}=\Gamma_{\;t\theta}^{\theta
}=\Gamma_{\;\theta t}^{\theta}=\Gamma_{\;t\varphi}^{\varphi}=\Gamma_{\;\varphi
t}^{\varphi}=\gamma(t),
\]
and
\[
\Gamma_{\;tt}^{t}=-\frac{\dot{\gamma}(t)}{\gamma(t)},\quad\Gamma_{\;rr}%
^{t}=\gamma(t),\quad\Gamma_{\;\theta\theta}^{t}=\gamma(t)r^{2},\quad
\Gamma_{\;\varphi\varphi}^{t}=\gamma(t)r^{2}\sin^{2}\theta.
\]

The non-metricity scalars for each connection are calculated%
\begin{equation}
Q_{1}\left(  \Gamma_{1}\right)  =-6H^{2}%
\end{equation}%
\begin{equation}
Q_{2}\left(  \Gamma_{2}\right)  =-6H^{2}+\frac{3}{a^{3}N}\left(  \frac
{a^{3}\gamma}{N}\right)  ^{\cdot}%
\end{equation}
and%
\begin{equation}
Q_{3}\left(  \Gamma_{3}\right)  =-6H^{2}+\frac{3}{a^{3}N}\left(
aN\gamma\right)  ^{\cdot}.
\end{equation}

Connection $\Gamma_{1}$ is the one defined in the coincidence gauge, while
connections $\Gamma_{2}$ and $\Gamma_{3}$ are defined in the non-coincidence
gauge. We observe that connection $\Gamma_{k}$ in the limit $k=0,$ reduces to
that of $\Gamma_{3}$, that is,~$\Gamma_{k}\left(  k\rightarrow0\right)
=\Gamma_{3}$ and $Q_{k}\left(  k\rightarrow0\right)  =Q_{3}\left(  \Gamma
_{3}\right)  $.

We proceed with the derivation of the minisuperspace Lagrangian and the field
equations for each family of connections in $f\left(  Q,B\right)  $-theory of gravity.

\subsubsection{Connection $\Gamma_{1}$}

For connection $\Gamma_{1}$, and the Ricciscalar (\ref{bd.002}) for the
spatially flat FLRW geometry we derive the boundary term%
\begin{equation}
B_{1}=B\left(  \Gamma_{1}\right)  =3\left(  6H^{2}+\frac{2}{N}\dot{H}\right)
.
\end{equation}

For the coincidence gauge, scalars $Q_{1}$ and $B_{1}$ have the same
functional form with the torsion scalar $T$ and the boundary $B_{T}$. As a
result $f\left(  Q,B\right)  -$gravity for the connection $\Gamma_{1}$ is
equivalent with the teleparallel $f\left(  T,B_{T}\right)  $-gravity.

The Lagrangian of the field equations is
\begin{equation}
L\left(  N,a,\dot{a},Q,B,\dot{B}\right)  =-\frac{6}{N}f_{,Q}a\dot{a}^{2}%
-\frac{6}{N}a^{2}\dot{f}_{,B}\dot{a}+Na^{3}\left(  f-f_{,Q}-f_{,B}\right)
\end{equation}
or equivalently in scalar field description%
\begin{equation}
L\left(  N,a,\dot{a},\phi,\zeta,\dot{\zeta}\right)  =-\frac{6}{N}\phi a\dot
{a}^{2}-\frac{6}{N}a^{2}\dot{a}\dot{\zeta}+Na^{3}V\left(  \phi,\zeta\right)  .
\label{bg.01}%
\end{equation}
where similarly as before $\phi=f_{,Q}$, $\zeta=f_{,B}$ and $V\left(
\phi,\zeta\right)  =\left(  f-f_{,Q}-f_{,B}\right)  $.

For the Lagrangian (\ref{bg.01}) in the three-dimensional space $\left\{
a,\phi,\zeta\right\}  $ we derive $\left\vert \frac{\partial^{2}L}{\partial
q\partial q}\right\vert =0$. Hence, the field equations are of fourth-order.

We select the constant lapse function $N=1$, and we derive the field equations%
\begin{equation}
0=6H\left(  \phi H+\dot{\zeta}\right)  +V\left(  \phi,\zeta\right)  ,
\end{equation}%
\begin{equation}
0=2\phi\left(  2\dot{H}+3H^{2}\right)  +4H\dot{\phi}+2\ddot{\zeta}+V\left(
\phi,\zeta\right)  ,
\end{equation}%
\begin{equation}
0=6H^{2}-V_{,\phi},
\end{equation}%
\begin{equation}
0=\dot{H}+3H^{2}+\frac{1}{6}V_{,\zeta}. \label{bg.05}%
\end{equation}

Modified Friedmann's equations are written in the equivalent form%
\begin{align*}
3H^{2}  &  =G_{eff}\rho_{f\left(  Q,B\right)  }^{\Gamma_{1}},\\
-2\dot{H}-3H^{2}  &  =G_{eff}p_{f\left(  Q,B\right)  }^{\Gamma_{1}},
\end{align*}
with energy density$~\rho_{f\left(  Q,B\right)  }^{\Gamma_{1}}$ and pressure
$p_{f\left(  Q,B\right)  }^{\Gamma_{1}}$ for the effective fluid
\begin{align}
\rho_{f\left(  Q,B\right)  }^{\Gamma_{1}}  &  =-\left(  3\dot{\zeta}H+\frac
{1}{2}V\left(  \phi,\zeta\right)  \right)  ,\\
p_{f\left(  Q,B\right)  }^{\Gamma_{1}}  &  =\left(  2H\dot{\phi}+\ddot{\zeta
}+\frac{1}{2}V\left(  \phi,\zeta\right)  \right)  .
\end{align}
and $G_{eff}=\frac{1}{\phi}$.

\subsubsection{Connection $\Gamma_{2}$}

For the non-coincidence connection $\Gamma_{2}$ it follows the boundary term%
\begin{equation}
B_{2}=B\left(  \Gamma_{2}\right)  =3\left(  6H^{2}+\frac{2}{N}\dot{H}-\frac
{3}{a^{3}N}\left(  \frac{a^{3}\gamma}{N}\right)  ^{\cdot}\right)  .
\end{equation}

Hence, by introducing Lagrange multipliers as we did for the generic
connection $\Gamma_{k}$, we determine the point-like Lagrangian
\begin{equation}
L\left(  N,a,\dot{a},Q,\dot{Q},B,\dot{B},\psi,\dot{\psi}\right)  =-\frac{6}%
{N}f_{,Q}a\dot{a}^{2}-\frac{6}{N}a^{2}\dot{f}_{,B}\dot{a}+3\left(  \dot
{f}_{,Q}-\dot{f}_{,B}\right)  \frac{a^{3}\dot{\psi}}{N}+Na^{3}\left(  f\left(
Q,B\right)  -f_{,Q}-f_{,B}\right)  ,
\end{equation}
or equivalently%
\begin{equation}
L\left(  N,a,\dot{a},\phi,\dot{\phi},\zeta,\dot{\zeta},\psi,\dot{\psi}\right)
=-\frac{6}{N}\phi a\dot{a}^{2}-\frac{6}{N}a^{2}\dot{a}\dot{\zeta}+3\left(
\dot{\phi}-\dot{\zeta}\right)  \frac{a^{3}\dot{\psi}}{N}+Na^{3}V\left(
\phi,\zeta\right)  ,
\end{equation}
in which $\gamma=\dot{\psi}$, $\phi=f_{,Q}$, $\zeta=f_{,B}$ and $V\left(
\phi,\zeta\right)  =\left(  f-f_{,Q}-f_{,B}\right)  $.

The field equations are of eight-order described by three scalar fields. For
$N=1$, the equations of motions for the scale factor and the three scalar
fields are%
\begin{equation}
0=6H\left(  \phi H+\dot{\zeta}\right)  -3\left(  \dot{\phi}-\dot{\zeta
}\right)  \dot{\psi}+NV\left(  \phi,\zeta\right)  ,
\end{equation}%
\begin{equation}
0=2\phi\left(  2\dot{H}+3H^{2}\right)  +4H\dot{\phi}-3\left(  \dot{\phi}%
-\dot{\zeta}\right)  \dot{\psi}+2\ddot{\zeta}+V\left(  \phi,\zeta\right)  ,
\end{equation}%
\begin{equation}
0=3\ddot{\psi}+9H\dot{\psi}-6H^{2}+V_{,\phi}%
\end{equation}%
\[
0=6\dot{H}+18H^{2}-9H\dot{\psi}-3\ddot{\psi}+V_{,\zeta}%
\]%
\begin{equation}
0=\ddot{\phi}-\ddot{\zeta}+3H\left(  \dot{\phi}-\dot{\zeta}\right)  .
\end{equation}

Hence. the effective geometric fluid has the following energy density and
pressure components%
\begin{align}
\rho_{f\left(  Q,B\right)  }^{\Gamma_{2}}  &  =-\left(  3\dot{\zeta}H+\frac
{3}{2}\left(  \dot{\phi}-\dot{\zeta}\right)  \dot{\psi}+\frac{1}{2}V\left(
\phi,\zeta\right)  \right)  ,\\
p_{f\left(  Q,B\right)  }^{\Gamma_{2}}  &  =\left(  2H\dot{\phi}-\frac{3}%
{2}\left(  \dot{\phi}-\dot{\zeta}\right)  \dot{\psi}+\ddot{\zeta}+\frac{1}%
{2}V\left(  \phi,\zeta\right)  \right)  ,
\end{align}
such that%
\begin{align*}
3H^{2}  &  =G_{eff}\rho_{f\left(  Q,B\right)  }^{\Gamma_{2}}\\
-2\dot{H}-3H^{2}  &  =G_{eff}p_{f\left(  Q,B\right)  }^{\Gamma_{2}},
\end{align*}
and $G_{eff}=\frac{1}{\phi}$.

\subsubsection{Connection $\Gamma_{3}$}

We set $k=0$ in (\ref{lan.01}), thus, the minisuperspace Lagrangian function
is%
\begin{equation}
L\left(  N,a,\dot{a},\phi,\dot{\phi},\zeta,\dot{\zeta},\Psi,\dot{\Psi}\right)
=-\frac{6}{N}\phi a\dot{a}^{2}-\frac{6}{N}a^{2}\dot{a}\dot{\zeta}%
-3Na\frac{\left(  \dot{\phi}-\dot{\zeta}\right)  }{\dot{\Psi}}+Na^{3}V\left(
\phi,\zeta\right)  .
\end{equation}
in which $\phi=f_{,Q}$, $\zeta=f_{,B}$ and $V\left(  \phi,\zeta\right)
=\left(  f-f_{,Q}-f_{,B}\right)  $, and the field equations are%
\begin{equation}
0=3\phi H^{2}+3H\dot{\zeta}-\frac{3}{2}\left(  \dot{\phi}-\dot{\zeta}\right)
\left(  \frac{1}{a^{2}\dot{\Psi}}\right)  +\frac{1}{2}V\left(  \phi
,\zeta\right)  ,
\end{equation}%
\begin{equation}
0=\dot{H}+\frac{3}{2}H^{2}+\frac{V}{4\phi}+H\frac{\dot{\phi}}{\phi}+\frac
{\dot{\zeta}-\dot{\phi}}{\phi}\left(  \frac{1}{4a^{2}\dot{\Psi}}\right)
+\frac{1}{2}\ddot{\zeta},
\end{equation}%
\begin{equation}
0=3\frac{\ddot{\Psi}}{\Psi}-\left(  3H-a^{2}\dot{\Psi}\left(  6H^{2}%
+9kH\dot{\Psi}-V_{,\phi}\right)  \right)  ,
\end{equation}%
\begin{equation}
0=6\dot{H}+18H^{2}+V_{,\zeta}-\frac{3}{a^{2}\dot{\Psi}^{2}}\left(  H\dot{\Psi
}-\ddot{\Psi}\right)  ,
\end{equation}%
\begin{equation}
0=3a\dot{\Psi}\left(  \ddot{\zeta}-\ddot{\phi}\right)  +a\left(  \dot{\zeta
}-\dot{\phi}\right)  \left(  \dot{\Psi}H-2\ddot{\Psi}\right)  .
\end{equation}

We conclude that the field equations are of eighth-order. Furthermore, the
geometric fluid has the energy density and pressure given by expressions
(\ref{lan.04}) and (\ref{lan.05}).

\section{$f\left(  Q,B\right)  =Q+F\left(  B\right)  $-Cosmology}

\label{sec5}

Based on the above analysis, we observe that $f\left(  Q,B\right)  $-theory
introduces a varying parameter $G_{eff}=\frac{1}{\phi}$,~$\phi=f_{,Q}$. Of
special interest are the $f\left(  Q,B\right)  =Q+F\left(  B\right)  $ models,
in which $G_{eff}=const$. and $\phi$ is always a constant. In this theory the
nonlinear terms of the Action Integral follow correspond to boundary
corrections. This approach has been previously explored in teleparallel
$f(T,B_{T})$-gravity, yielding numerous interesting results. Specifically,
$f(T,B_{T}) = T + F(B_{T})$ has the potential to explain both the late and
early-time acceleration phases of the universe \cite{ftb5}.

In the $f(Q,B) = Q + F(B)$ theory, the number of field equations is reduced by
one, as $\phi$ is not a dynamic parameter but a constant, i.e., $\phi=1$.
Below, we provide the sets of field equations in $f(Q,B) = Q + F(B)$-theory
for the four different families of connections.

\subsection{Connection $\Gamma_{1}$}

For the connection $\Gamma_{1}$ defined in the coincidence gauge, the
minisuperspace Lagrangian is
\begin{equation}
L\left(  N,a,\dot{a},\phi,\zeta,\dot{\zeta}\right)  =-\frac{6}{N}a\dot{a}%
^{2}-\frac{6}{N}a^{2}\dot{a}\dot{\zeta}+Na^{3}V\left(  \zeta\right)  .
\end{equation}
where $\zeta=f_{,B}$ and $V\left(  \zeta\right)  =\left(  f-f_{,B}\right)  $.

Thus, for $N=1$, the field equations are%
\begin{align}
3H^{2}  &  =\rho_{f\left(  Q,B\right)  }^{\Gamma_{1}},\\
-2\dot{H}-3H^{2}  &  =p_{f\left(  Q,B\right)  }^{\Gamma_{1}},
\end{align}
in which
\begin{align}
\rho_{f\left(  Q,B\right)  }^{\Gamma_{1}}  &  =-\left(  3\dot{\zeta}H+\frac
{1}{2}V\left(  \zeta\right)  \right)  ,\\
p_{f\left(  Q,B\right)  }^{\Gamma_{1}}  &  =\left(  \ddot{\zeta}+\frac{1}%
{2}V\left(  \zeta\right)  \right)  ,
\end{align}
and the scalar field $\zeta$ satisfies the equation of motion (\ref{bg.05}).
Because the theory is equivalent to the teleparallel $f\left(  T,B_{T}\right)
=T+F\left(  B_{T}\right)  $ model, the results of the latter theory are valid
for the the symmetric teleparallel theory with boundary term.

\subsection{Connection $\Gamma_{2}$}

For connection $\Gamma_{2}$ defined in the non-coincidence gauge, the
minisuperspace Lagrangian reads
\begin{equation}
L\left(  N,a,\dot{a},\phi,\dot{\phi},\zeta,\dot{\zeta},\psi,\dot{\psi}\right)
=-\frac{6}{N}a\dot{a}^{2}-\frac{6}{N}a^{2}\dot{a}\dot{\zeta}-3a^{3}\frac
{\dot{\zeta}\dot{\psi}}{N}+Na^{3}V\left(  \zeta\right)  , \label{con.021}%
\end{equation}
in which $\gamma=\dot{\psi}$, $\zeta=f_{,B}$ and $V\left(  \phi,\zeta\right)
=\left(  f-f_{,B}\right)  $.

For $N=1$, modified Friedmann's equations are%
\begin{align}
3H^{2}  &  =\rho_{f\left(  Q,B\right)  }^{\Gamma_{2}}\label{con.022}\\
-2\dot{H}-3H^{2}  &  =p_{f\left(  Q,B\right)  }^{\Gamma_{2}}, \label{con.023}%
\end{align}
with
\begin{align}
\rho_{f\left(  Q,B\right)  }^{\Gamma_{2}}  &  =-\left(  3\dot{\zeta}H-\frac
{3}{2}\dot{\zeta}\dot{\psi}+\frac{1}{2}V\left(  \zeta\right)  \right)
,\label{con.024}\\
p_{f\left(  Q,B\right)  }^{\Gamma_{2}}  &  =\left(  \frac{3}{2}\dot{\zeta}%
\dot{\psi}+\ddot{\zeta}+\frac{1}{2}V\left(  \zeta\right)  \right)  ,
\label{con.025}%
\end{align}
where the scalar fields $\zeta$ and $\psi$ satisfy the field equations%
\begin{align}
0  &  =6\dot{H}+18H^{2}-9H\dot{\psi}-3\ddot{\psi}+V_{,\zeta}\label{con.027}\\
0  &  =\ddot{\zeta}+3H\dot{\zeta}. \label{con.028}%
\end{align}

\subsection{Connection $\Gamma_{3}$}

For the connection $\Gamma_{3}$ the minisuperspace Lagrangian becomes%

\begin{equation}
L\left(  N,a,\dot{a},\phi,\dot{\phi},\zeta,\dot{\zeta},\Psi,\dot{\Psi}\right)
=-\frac{6}{N}a\dot{a}^{2}-\frac{6}{N}a^{2}\dot{a}\dot{\zeta}+3Na\frac
{\dot{\zeta}}{\dot{\Psi}}+Na^{3}V\left(  \zeta\right)  .
\end{equation}
with $\gamma=\frac{1}{\dot{\Psi}}$,, $\zeta=f_{,B}$ and $V\left(
\zeta\right)  =\left(  f-f_{,B}\right)  $.

Furthermore, for $N=1$, the gravitational field equations are%
\begin{align}
3H^{2}  &  =\rho_{f\left(  Q,B\right)  }^{\Gamma_{3}}\\
-2\dot{H}-3H^{2}  &  =p_{f\left(  Q,B\right)  }^{\Gamma_{3}},
\end{align}
where
\begin{align}
\rho_{f\left(  Q,B\right)  }^{\Gamma_{k}}  &  =-\left(  3H\dot{\zeta}+\frac
{3}{2}\frac{\dot{\zeta}}{a^{2}\dot{\Psi}}+\frac{1}{2}V\left(  \zeta\right)
\right)  ,\\
p_{f\left(  Q,B\right)  }^{\Gamma_{k}}  &  =\frac{V}{2}+\frac{\dot{\zeta}%
}{8a^{2}\dot{\Psi}}.
\end{align}%
\begin{equation}
0=3\phi H^{2}+3H\dot{\zeta}-\frac{3}{2}\left(  \dot{\phi}-\dot{\zeta}\right)
\left(  \frac{1}{a^{2}\dot{\Psi}}\right)  +\frac{1}{2}V\left(  \phi
,\zeta\right)  ,
\end{equation}
where the scalar fields $\zeta$ and $\Psi$ satisfy the equations of motion%
\begin{equation}
0=6\dot{H}+18H^{2}+V_{,\zeta}-\frac{3}{a^{2}\dot{\Psi}^{2}}\left(  H\dot{\Psi
}-\ddot{\Psi}\right)  ,
\end{equation}%
\begin{equation}
0=3\dot{\Psi}\ddot{\zeta}+\dot{\zeta}\left(  \dot{\Psi}H-2\ddot{\Psi}\right)
.
\end{equation}

\subsection{Connection $\Gamma_{k}$}

Finally, for the fourth connection $\Gamma_{k}$ where curvature is nonzero,
the minisuperspace Lagrangian is written
\begin{equation}
L\left(  N,a,\dot{a},\phi,\dot{\phi},\zeta,\dot{\zeta},\Psi,\dot{\Psi}\right)
=-\frac{6}{N}a\dot{a}^{2}+6Nak-\frac{6}{N}a^{2}\dot{a}\dot{\zeta}+3\dot{\zeta
}\left(  a\frac{N}{\dot{\Psi}}-k\frac{a^{3}}{N}\dot{\Psi}\right)
+Na^{3}V\left(  \zeta\right)  .
\end{equation}
in which $\gamma=\frac{1}{\dot{\Psi}}$,, $\zeta=f_{,B}$ and $V\left(
\zeta\right)  =\left(  f-f_{,B}\right)  $.

The gravitational field equations in the presence of curvature are%
\begin{align}
3\left(  H^{2}+\frac{k}{a^{2}}\right)   &  =G_{eff}\rho_{f\left(  Q,B\right)
}^{\Gamma_{k}},\\
-2\dot{H}-3H-\frac{k}{a^{2}}  &  =G_{eff}p_{f\left(  Q,B\right)  }^{\Gamma
_{k}}.
\end{align}

The energy density and pressure components of the geometric fluid are defined
as%
\begin{align}
\rho_{f\left(  Q,B\right)  }^{\Gamma_{k}}  &  =-\left(  3H\dot{\zeta}+\frac
{3}{2}\dot{\zeta}\left(  \frac{1}{a^{2}\dot{\Psi}}+k\dot{\Psi}\right)
+\frac{1}{2}V\left(  \zeta\right)  \right)  ,\\
p_{f\left(  Q,B\right)  }^{\Gamma_{k}}  &  =\frac{V}{2}+\frac{\dot{\zeta}}%
{2}\left(  \frac{1}{4a^{2}\dot{\Psi}}-\frac{3k}{4}\dot{\Psi}\right)  .
\end{align}
For the scalar fields we derive the equations of motion
\begin{equation}
0=6\dot{H}+9H\left(  2H+k\dot{\Psi}\right)  +3k\ddot{\Psi}+V_{,\zeta}-\frac
{3}{a^{2}\dot{\Psi}^{2}}\left(  H\dot{\Psi}-\ddot{\Psi}\right)  ,
\end{equation}%
\begin{equation}
0=3a\dot{\Psi}\left(  1+a^{2}k\dot{\Psi}^{2}\right)  \ddot{\zeta}+\dot{\zeta
}\left(  \dot{\Psi}H\left(  1+3a^{2}k\dot{\Psi}^{2}\right)  -2\ddot{\Psi
}\right)  .
\end{equation}

From the above results we remark that in $f\left(  Q,B\right)  =Q+F\left(
B\right)  $ gravity the field equations are of fourth-order in the coincidence
gauge and of sixth-order in the non-coincidence gauge.

\section{New solution in the non-coincidence gauge}

\label{sec6}

We focus on the field equations (\ref{con.022})-(\ref{con.028}) for the
connection $\Gamma_{2}$ in the case of $f\left(  Q,B\right)  =Q+F\left(
B\right)  $ theory. From equation (\ref{con.028}) we construct the
conservation law
\begin{equation}
I_{0}=a^{3}\dot{\zeta}.
\end{equation}

For the exponential potential $V\left(  \zeta\right)  =V_{0}e^{\lambda\zeta}%
$,~i.e. $F\left(  B\right)  =-\frac{B}{\lambda}\ln\left(  -\frac{B}{\lambda
V_{0}}\right)  -\frac{B}{\lambda}$; we are able to write the second
conservation law%
\begin{equation}
I_{1}=a^{2}\left(  2\left(  \lambda-3\right)  \dot{a}+a\left(  \lambda
\dot{\zeta}-3\dot{\psi}\right)  \right)  .
\end{equation}
\newline In order to determine the latter conservation law we applied the
method of variational symmetries which has been widely used in modified
theories of gravity, for more details we refer the reader to \cite{nos1,nos2}
and references therein.

With the use of the two conservation laws and of the constraint equation
(\ref{con.022}), the second Friedmann equation reads%
\begin{equation}
2a^{5}\ddot{a}-2a^{2}\dot{a}\left(  I_{0}\lambda+a^{2}\dot{a}\right)
-I_{0}\left(  I_{0}\lambda-I_{1}\right)  =0. \label{so.01}%
\end{equation}

For $I_{0}\lambda-I_{1}=0$, we are able to determine the closed form solution%
\begin{equation}
a\left(  t\right)  =\left(  a_{1}e^{a_{0}t}+I_{0}\lambda\right)  ^{\frac{1}%
{3}}.
\end{equation}

Consequently, the Hubble function is derived%
\begin{equation}
H^{2}\left(  a\right)  =\sqrt{\left(  \frac{a_{0}}{3}\right)  ^{2}%
-\frac{2a_{0}I_{0}\lambda}{9}a^{-3}+\frac{\left(  I_{0}\lambda\right)  ^{2}%
}{9}a^{-6}}.
\end{equation}
This analytic solution describes a universe with a cosmological constant, dark
matter and a stiff fluid. Indeed when $\frac{\left(  I_{0}\lambda\right)
^{2}}{9}a^{-6}$ is neglected, i.e. $\frac{\left(  I_{0}\lambda\right)  ^{2}%
}{9}a^{-6}\rightarrow0$ , the limit of $\Lambda$CDM universe is recovered.
Furthermore, we calculate the deceleration parameter%
\[
q\left(  a\right)  =2-\frac{3a^{3}}{a^{3}-I_{0}\lambda}\text{,}%
\]
from where it follows that for large values of $a$, $q\left(  a\right)
\rightarrow-1$, that is, the de Sitter universe is recovered. Acceleration
point is occurred when $2<\frac{3a^{3}}{a^{3}-I_{0}\lambda}$.

In order to solve equation\ (\ref{so.01}) we apply the Lie symmetry analysis
\cite{nos1}. We find that equation (\ref{so.01}) is invariant under the action
of the elements of a two-dimensional Lie algebra consisted by the vector
fields $X_{1}=\partial_{t}$ and $X_{2}=3t\partial_{t}+a\partial_{a}$. \ The
application of the Lie invariants of $X_{2}$ indicate the existence of the
exact solution $a\left(  t\right)  =\bar{a}_{0}t^{\frac{1}{3}}$ with
constraint equation $2a_{0}^{3}\left(  a_{0}^{3}+\lambda I_{0}\right)
=3I_{0}\left(  I_{0}\lambda-I_{1}\right)  $.

On the other hand, the application of the Lie symmetry vector $X_{1}$ provides
the reduced equation%
\begin{equation}
\frac{dA}{da}=\frac{I_{0}\left(  I_{1}-I_{0}\lambda\right)  }{2a^{5}}%
A^{3}-\frac{I_{0}\lambda}{a^{3}}A^{2}-\frac{1}{a}A~,~A\left(  t\right)
=\frac{1}{\dot{a}}~.
\end{equation}
This is an Abel type equation.

\begin{figure}[ptb]
\centering\includegraphics[width=1\textwidth]{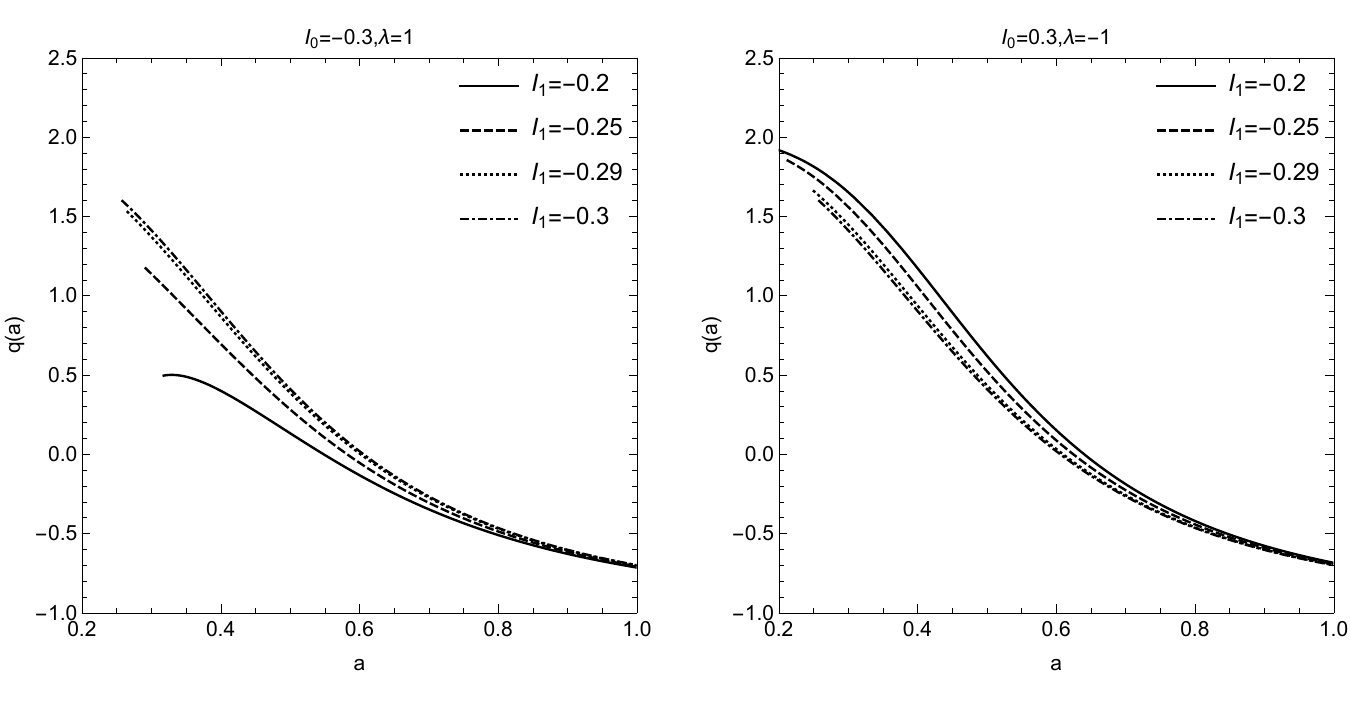}\caption{Qualitative
evolution for the effective deceleration parameter $q=-1-\frac{\dot{H}}{H^{2}%
}~$as it is given by the numerical solution of equation (\ref{so.01}). The
plots are for different values of the free parameters.}%
\label{fig1}%
\end{figure}

\begin{figure}[ptb]
\centering\includegraphics[width=1\textwidth]{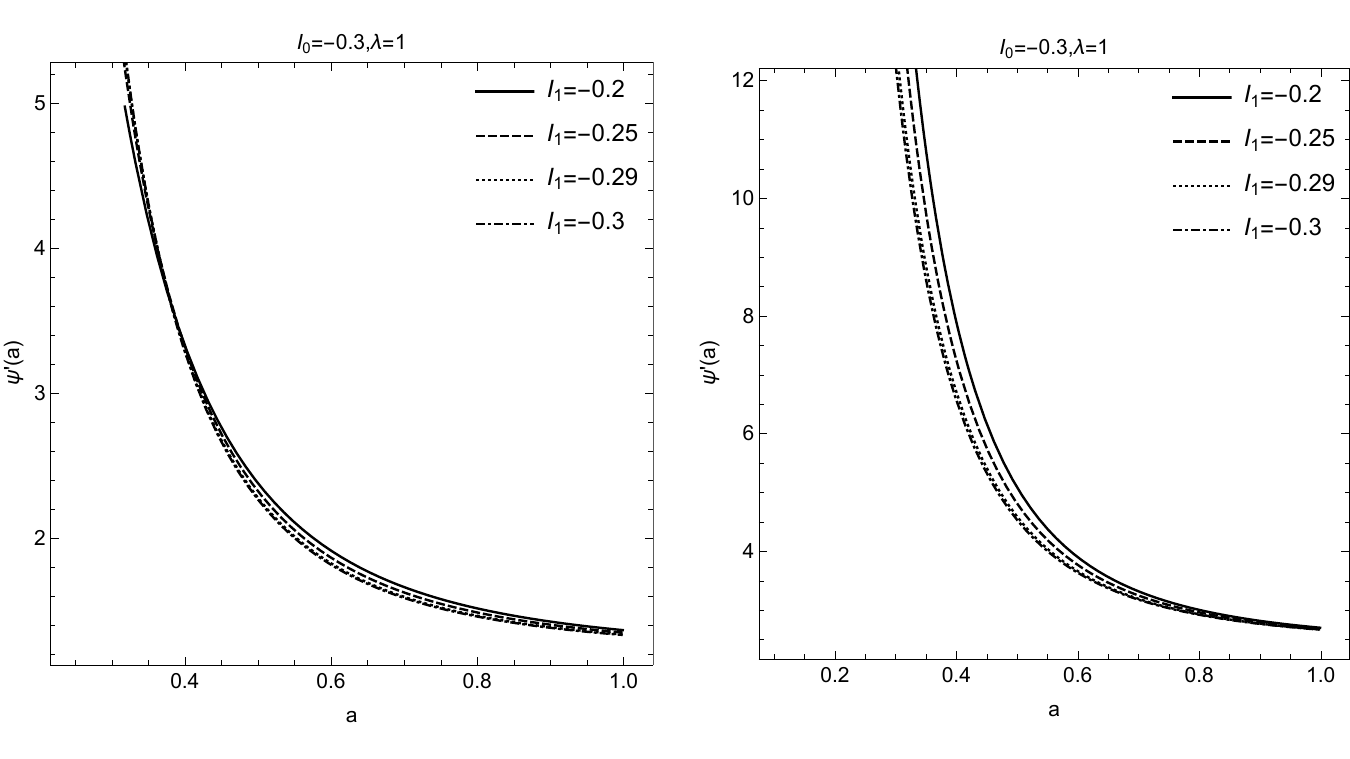}\caption{Qualitative
evolution for the function $\gamma=\dot{\psi}~$as it is given by the numerical
solution of equation (\ref{so.01}). The plots are for different values of the
free parameters. We observe that as the solution reaches the de Sitter
universe connection $\Gamma_{2}$ reach to the limit of $\Gamma_{1}$.}%
\label{fig2}%
\end{figure}

In Figs. \ref{fig1} and \ref{fig2} we present the qualitative evolution of the
deceleration parameter $q=-1-\frac{\dot{H}}{H^{2}}$ and of the function
$\gamma=\dot{\psi}$, as it is given after the numerical simulation of equation
(\ref{so.01}). We observe that the de Sitter universe is a future attractor
for the cosmological model, and in the limit of the de Sitter universe
connection $\Gamma_{2}$ takes the form of $\Gamma_{1}$. 

\section{Conclusions}

\label{sec7}

We conducted a study on FLRW cosmology within the framework of symmetric
teleparallel theory, considering boundary corrections in the gravitational
Lagrangian. In this context, and for the four different families of
connections, we determined the minisuperspace description of the field
equations. To achieve this, we introduced Lagrange multipliers, enabling us to
express the higher-order derivatives of the field equations in terms of scalar
fields. As a result, we were able to recast the cosmological field equations
into the equivalent form of multi-scalar field cosmology. In the case of
connections defined in the non-coincidence gauge, $f(Q,B)$-gravity is
characterized by three scalar fields. However, in the limiting case of the
$f(Q,B) = Q + F(B)$ model, the field equations are described by two scalar
fields. Conversely, for the connection defined in the coincidence gauge, there
exists only one scalar field, and the field equations are of fourth-order.
It's worth noting that $f(Q)$-gravity introduces two scalar fields when the
connection is defined in the non-coincidence gauge.

This scalar field description and the derivation of the minisuperspace
representation are crucial for further investigations into the dynamic
evolution of physical variables within the theory. Moreover, the
minisuperspace Lagrangian can be employed to establish the Hamiltonian
formalism of the model and derive the Wheeler-DeWitt equation of quantum cosmology.

To illustrate the practical application of the minisuperspace description, we
employed the method of variational symmetries and successfully determined an
integrable cosmological model. We were able to express the analytic solution
in terms of the Abel equation. This particular cosmological model not only
accounts for cosmic acceleration but also includes a dark matter component in
the Hubble function.

These results suggest that $f(Q,B)$-theory holds promise as a viable
cosmological framework. However, one notable implication is the significant
increase in degrees of freedom introduced by this theory. Therefore, the new
scalar fields must be capable of describing a wide range of cosmological
phenomena. In future work, we plan to investigate whether boundary correction
terms in symmetric teleparallel theory can resolve cosmological tensions and
whether the theory can provide explanations for eras in the cosmological
history beyond late-time acceleration.

\textbf{Data Availability Statements:} Data sharing is not applicable to this
article as no datasets were generated or analyzed during the current study.

\begin{acknowledgments}
The author thanks the support of Vicerrector\'{\i}a de Investigaci\'{o}n y
Desarrollo Tecnol\'{o}gico (Vridt) at Universidad Cat\'{o}lica del Norte
through N\'{u}cleo de Investigaci\'{o}n Geometr\'{\i}a Diferencial y
Aplicaciones, Resoluci\'{o}n Vridt No - 098/2022.
\end{acknowledgments}

\end{document}